\RequirePackage{fix-cm}
\documentclass[twocolumn,epjc3]{svjour3}  
\smartqed  
\RequirePackage{graphicx}
\RequirePackage{float}
\RequirePackage{hyperref}
\RequirePackage{amsmath}
\usepackage{mwe}
\usepackage{widetext}
\RequirePackage{amssymb}
\RequirePackage{mathtools}   
\usepackage{orcidlink}
\usepackage{booktabs}
\usepackage{caption}
\usepackage{multirow}
\usepackage{dcolumn}
\usepackage[caption=false]{subfig}
\usepackage{float}   
\journalname{Eur. Phys. J. C}
\begin{document}

\title{Generalised Ellis-Bronnikov Wormholes in $f(R)$ Gravity}


\author{Oleksii Sokoliuk\thanksref{e1,addr1,addr2}
        \and
        Sanjay Mandal\thanksref{e2,addr3}
        \and
        P.K. Sahoo\thanksref{e3,addr3}
        \and Alexander Baransky\thanksref{e4,addr2}}

\thankstext{e1}{e-mail: oleksii.sokoliuk@mao.kiev.ua}
\thankstext{e2}{e-mail: sanjaymandal960@gmail.com}
\thankstext{e3}{e-mail: pksahoo@hyderabad.bits-pilani.ac.in}
\thankstext{e4}{e-mail: abransky@ukr.net}


\institute{Main Astronomical Observatory of the NAS of Ukraine (MAO NASU),Kyiv, 03143, Ukraine \label{addr1} \and Astronomical Observatory, Taras Shevchenko National University of Kyiv, 3 Observatorna St., 04053 Kyiv, Ukraine \label{addr2} \and Department of Mathematics, Birla Institute of
Technology and Science-Pilani, Hyderabad Campus, Hyderabad-500078,
India \label{addr3}}

\date{Received: 25 Jan. 2022 / Accepted: 23 March 2022}

\maketitle

\begin{abstract}
In this manuscript, we construct generalized Ellis-Bronnikov wormholes in the context of $f(R)$ modified theories of gravity. We consider that the matter driving the wormhole satisfies the energy conditions so that it is the effective energy-momentum tensor containing the higher-order derivatives of curvature terms that violate the null energy condition. Thus, the gravitational fluid is interpreted by the higher-order derivatives of curvature terms to represent the wormhole geometries and is fundamentally different from its counter representation in general relativity. In particular, we explore the wormhole geometries by presuming various well-known forms of Lagrangian $f(R)$. In addition, for the seek of completeness, we discuss modified Tolman-Oppenheimer-Volkov, volume integral quantifier, and total gravitational energy.
\end{abstract}

\section{Introduction}
Various independent observational measurements have confirmed that the universe is going through the accelerated expansion phase \cite{ref1,ref2,ref3,ref4}. Several proposals have been proposed in the literature to explain this phenomenon, ranging from dark energy models to the modified theories of gravity. Moreover, it is well-known that the general theory of relativity (GR) is quite successful in the fundamental basis. Also, it is able to describe the accelerated expansion process of the universe by introducing cosmological constant into the Einstein field equation. But, the cosmological constant leads to various misleading issues \cite{ref5,ref6}. Furthermore, the Einstein field equation of GR was derived by Hilbert using the action principle by considering Ricci scalar, $R$, in the Lagrangian gravitational density. However, there are no reasons to limit the gravitational Lagrangian to this form a priori, and numerous generalizations have been proposed. In particular, a more general modification of the gravitational Lagrangian was done by introducing a general function of scalar invariant $R$, called $f(R)$ gravity \cite{ref7} and further developed in \cite{ref8,ref9}.

In this view, a family of $f(R)$ gravity theories has been successfully examined in an attempt to describe the universe's late-time accelerated expansion \cite{ref10,ref11}. The inflationary scenario of the universe was motivated by the early development of $f(R)$ theories; as a result, some interesting forms of $f(R)$ were considered to explore the universe \cite{ref12}. Moreover, the accelerated expansion of the universe can be explained in the framework of $f(R)$ theory \cite{ref13}. Furthermore, the coupling between the arbitrary function of $R$ and matter Lagrangian density has been explored \cite{ref14,ref15,ref16,ref17,ref18}, and several viable conditions have been derived from testing the cosmological models in $f(R)$ gravity \cite{ref19,ref20,ref21,ref22,ref23,ref24,ref25}. In the Solar System test, most of the proposed cosmological $f(R)$ models have been ruled out so far \cite{ref26,ref27,ref28,ref29,ref30,ref31}, although suitable models do exist \cite{ref32,ref33,ref34,ref35,ref36}. In addition, $f(R)$ gravity can explore the galactic dynamics of the massive test particle without dark matter \cite{ref37,ref38,ref39}.

This study extends the analysis of static and spherically symmetric spacetime and explores the traversable wormhole geometries in a renaissance of $f(R)$ gravity. Wormholes are hypothetical tunnels connecting different spacetimes or two different regions of the same spacetime, and possible it helps the observer to go freely from one region to another. However, it is worth mentioning here that the wormhole solutions are primarily used as ``Gedanken experiments" and as a theoretician's view of the foundation of general relativity. In classical GR, wormholes are supported by the exotic form of matter to make it traversable. As a result, the stress-energy tensor of matter fluid violates the null energy condition (NEC) \cite{ref40}. Note that NEC is given by $T_{\mu\nu}k^{\mu}k^{\nu}\geq 0$, where $k^{\mu}$ is any null vector. Thus it is an important and open challenge in wormhole physics to find a viable matter source to support this exotic spacetime. In this regard, several candidates have been proposed in the literature, for instance, wormholes in Einstein-Gauss-Bonnet theory \cite{ref41}, static wormhole solution for higher-dimensional gravity in vacuum \cite{ref42}, wormholes on braneworld \cite{ref43,ref44,ref45,ref46}, thin accretion disks in wormhole geometries \cite{ref47}, wormhole geometries in $f(R) $modified theories of gravity \cite{ref48,ref49,ref50,ref52,ref52,ref53}, wormhole models in $f(R,T)$ gravity theory \cite{ref54,ref55}, and some recent studies (see the ref. \cite{ref56,ref57,ref58,ref59,ref60,ref61}).

Furthermore, energy conditions play a significant role in the matter profiles of spacetime structure. At the same time, NEC plays a crucial role in wormhole geometry. Recently, F. Rahaman and his research group studied the energy conditions for various wormhole structures by considering various shape functions; they have examined the energy conditions for the wormhole geometry in $f(Q)$ gravity \cite{ref61a}. Also, some recent studies see, e.g. \cite{ref61b}, where energy conditions, especially NEC, are tested to present an alternative explanation of exotic matter through modified theories of gravity. B. Narzilloev et al. investigated how one can distinguish the particle motion around a static axially symmetric wormhole from a black hole \cite{ref61c}. Chakraborty and Kar studied how NEC violation can avoid a zero proper volume end-state of a collapsing wormhole \cite{ref61d}. Karakasis et al. discussed the wormhole geometries by introducing a phantom scalar field in the gravitational action of $f(R)$ gravity; as a result, it becomes ghost-free and avoids the tachyonic instability \cite{ref61e}. Farook et al. studied shadows of a particular class of rotating wormhole, and they compose the null geodesics and study the effects of the parameters on the photon orbit \cite{ref61f}. Kuhfitting addressed two fundamental issues concerning Morris-Throne wormholes in the five-dimensional spacetime, such as the origin of exotic matter and frequently inexplicable enormous radial tension at the throat \cite{ref61g}. Conditions for safe travel through a thin-shell wormhole throat are analyzed in \cite{ref61h}. The above literature discussed the static symmetric spacetime, whereas the generalization of such metric has been explored mainly in general relativity. Therefore, we aim to explore the generalization of the static spherically symmetric spacetime such as GEB wormholes in the modified theory of gravity (in particular, $f(R)$ gravity).

This manuscript is organized in the following manner: Section \ref{sec2} discusses the construction of generalized Ellis-Bronnikov (GEB) wormholes in the background of $f(R)$ modified theories of gravity. Then, we present the energy conditions for the GEB wormholes in section \ref{sec3}. We discuss the three types of wormholes geometries and test their energy conditions. In addition, we discuss modified TOV equation (MTOV), volume integral quantifier (VIQ), and total gravitational energy in sections \ref{sec4}, \ref{sec5}, and \ref{sec6}, respectively. Finally, gathering all the outcomes from our study, we concluded in section \ref{sec7}.

\section{construction of Generalised Ellis-Bronnikov Wormhole}\label{sec2}

The line element of the Ellis-Bronnikov wormhole is given by

\begin{equation}
\label{1}
ds^2=-dt^2+\left(1-\frac{b_0^2}{r^2} \right)^{-1}dr^2+r^2 d\theta^2+r^2 sin^2(\theta)d\phi^2.
\end{equation}

The generalized Ellis-Bronnikov (GEB) spacetime can be written as

\begin{equation}
\label{2}
ds^2=-dt^2+dl^2+r^2(l)d\theta^2+r^2(l)sin^2(\theta)d\phi^2,
\end{equation}
where $r(l)=(b_0^2+l^m)^{1/m}$.

The parameter $m$ takes only even values to make $r(l)$ smooth over the entire domain of the so-called 'tortoise' or 'proper radial distance' coordinate $l$ (where $-\infty \leq l\leq \infty$) \textbf{and $b_0$ is the wormhole throat}. Metric \eqref{2}, in terms of usual radial coordinate $r$, can be written as
\begin{equation}
\label{3}
ds^2=-dt^2+\left(1-\frac{b(r)}{r} \right)^{-1}dr^2+r^2 d\theta^2+r^2 sin^2(\theta)d\phi^2,
\end{equation}
where $r$ and $l$ are related through the shape function $b(r)$ as,
\begin{equation}
\label{4}
dl^2=\frac{dr^2}{1-\frac{b(r)}{r}},
\end{equation}

\begin{equation}\label{5}
b(r)=r-r^{(3-2m)}\left(r^m-b_0^m\right)^{2-2/m}.
\end{equation}

The action in $f(R)$ gravity reads

\begin{equation}
\label{6}
\mathcal{S}=\frac{1}{2\kappa}\int \sqrt{-g}f(R)d^4x+\int \mathcal{L}_m
d^4x,
\end{equation}
where $\kappa=8\pi G, \mathcal{L}_m$ is the matter Lagrangian.

By varying this action with respect to metric we find

\begin{equation}
\label{7}
f_R(R)R_{\mu\nu}-\frac{1}{2}g_{\mu\nu} f(R)-(\nabla_{\mu}\nabla_{\nu}-g_{\mu\nu}\square)f_R(R)=\kappa T_{\mu\nu},
\end{equation}
where $f_R(R)=df(R)/dR$ and $T_{\mu\nu}$ is the energy momentum tensor of the matter which is defined by
\begin{align*}
T_{\mu\nu}=-\frac{2}{\sqrt{-g}}\frac{\delta\left(\sqrt{-g}\,\mathcal{L}_m\right)}{\delta g^{\mu\nu}}.
\end{align*}

Considering the contraction of Eq. \eqref{7}, provides the following relationship
\begin{equation}
\label{8}
R f_R(R)-2f(R)+3\square f_R(R)=T,
\end{equation}
where $R$ is the Ricci scalar, and $T=T^{\mu}_{\mu}$ is the stress of the energy-momentum tensor.

The trace equation \eqref{8} can be used to simplify the field
equations and then can be kept as a constraint equation.
Thus, substituting the trace equation into Eq. \eqref{7}, and
reorganizing the terms we end up with the following
gravitational field equation
\begin{equation}
\label{9}
G_{\mu\nu}=R_{\mu\nu}-\frac{1}{2}R g_{\mu\nu}=T^{eff}_{\mu\nu},
\end{equation}
where the effective stress-energy tensor is given by
\begin{multline}
\label{10}
T^{eff}_{\mu\nu}=\frac{1}{f_R(R)}\bigg[T_{\mu\nu}+\nabla_{\mu}\nabla_{\nu} f_R(R)\\
-\frac{1}{4}g_{\mu\nu}(Rf_R(R)+\square f_R(R)+T)\bigg].
\end{multline}
We consider the anisotropic energy momentum tensor for the matter distribution as
\begin{equation}
\label{11}
T_{\mu\nu}=(\rho+p_t)U_{\mu}U_{\nu}+p_tg_{\mu\nu}+(p_r-p_t)\chi_{\mu}\chi_{\nu},
\end{equation}
where $U^{\mu}$ is the four velocity, $\chi^{\mu}$ is the unit spacelike vector in the radial direction. $\rho$ is the energy density, $p_r$ is the radial pressure measured in the direction of $\chi^{\mu}$ and $p_t$ is the transverse pressure measured in the orthogonal direction to $\chi^{\mu}$. The stress of energy-momentum tensor for the above considerations can read the following:
\begin{align*}
T^{\mu}_{\nu}=\text{diag}[-\rho(l),p_r(l),p_t(l),p_t(l)].
\end{align*}
Also, we can write $T=-\rho+p_r+2p_t$.

Now, the effective field equation \eqref{9} gives the following relationships

\begin{equation}
\label{12}
-\frac{-1+r'(l)^2+2r(l)r''(l)}{r^2}=\frac{1}{f_R}\left[\rho+K(r)\right],
\end{equation}

\begin{equation}
\label{13}
\frac{-1+r'(l)^2}{r(l)^2}=\frac{1}{f_R}\left[p_r(l)+f_R''-K(r)\right],
\end{equation}

\begin{equation}
\label{14}
\frac{r''(l)}{r(l)}=\frac{1}{f_R}\left[p_t+\frac{r'(l)}{r(l)}f_R'-K(r)\right],
\end{equation}
where a prime (') denotes derivative with respect to $l$. The term $K(r)$ is defines as
\begin{equation}
\label{15}
K(r)=\frac{1}{4}(R\,f_R+\square f_R+T),
\end{equation}
for the notational simplicity. The curvature scalar $R$ is given by
\begin{equation}
\label{16}
R=\frac{2(1-r'(l)^2-2r(l)r''(l))}{r(l)^2},
\end{equation}

and $\square f_R$ is given by the following relation

\begin{equation}
\label{17}
\square f_R=f_R''+2\frac{r'(l)}{r(l)}f_R'.
\end{equation} 
Now, one can use this set up to explore various GEB wormholes in the formalism of $f(R)$ modifiesd theories of gravity.

\section{GEB wormhole energy conditions and viable $f(R)$ gravities}\label{sec3}

Energy conditions (further - EC's) violation is the main problem of cosmological wormholes. Energy conditions usually could tell us whether the fluid is physically realistic or not. EC's origin are temporal Raychaudhuri equations along with the requirement that gravity is attractive. Raychaudhuri equation reads \cite{ref11,ref62,ref63,ref64}:
\begin{equation}
    \frac{d\theta}{d\tau}=-\frac{1}{3}\theta^2-\sigma_{\mu\nu}\sigma^{\mu\nu}+\omega_{\mu\nu}\omega^{\mu\nu}-R_{\mu\nu}u^\mu u^\nu
\end{equation}
Where $u^\nu$ is the timelike geodesics, $\theta$, $\sigma_{\mu\nu}$ and $\omega_{\mu\nu}$ are expansion, shear and rotation associated with the vector field $u^\mu$, $R_{\mu\nu}$ is the regular Ricci tensor. Also, for the null-like vector field $\eta^\mu$ Raychaudhuri equation has the following form:
\begin{equation}
        \frac{d\theta}{d\tau}=-\frac{1}{2}\theta^2-\sigma_{\mu\nu}\sigma^{\mu\nu}+\omega_{\mu\nu}\omega^{\mu\nu}-R_{\mu\nu}\eta^\mu \eta^\nu
\end{equation}
As we already stated, we assume that gravity nature is attractive and then $\theta<0$. For that case, timelike and null-like Raychaudhuri equations satisfy:
\begin{equation}
    R_{\mu\nu}u^\mu u^\nu\geq0
    \label{eq:18}
\end{equation}
\begin{equation}
    R_{\mu\nu}\eta^\mu \eta^\nu\geq0
    \label{eq:19}
\end{equation}
In the current paper we are going to test the different energy conditions in our $f(R)$ models for GEB traversable wormholes. Assuming the anisotropic matter distribution and following the methodology in \cite{ref48}, the energy conditions under this framework reads:
\begin{itemize}
    \item Null Energy Condition (NEC): $\rho^{eff}+p_r^{eff} \geq 0\land \rho^{eff}+p_t^{eff} \geq 0$
    \item Weak Energy Condition (WEC): $\rho^{eff}\geq0$ and $\rho^{eff}+p_r^{eff} \geq 0 \land \rho^{eff}+p_t^{eff} \geq 0$
    \item Strong Energy Condition (SEC): $\rho^{eff} + p_r^{eff} + 2p_t^{eff} \geq 0$
    \item Dominant Energy Condition (DEC): $\rho^{eff} \geq |p_r^{eff}| \land \rho^{eff} \geq |p_t^{eff}|$
\end{itemize}
Here,
\begin{equation}
\label{12a}
\rho^{eff}=\frac{1}{f_R}\left[\rho+K(r)\right],
\end{equation}

\begin{equation}
\label{13a}
p_r^{eff}=\frac{1}{f_R}\left[p_r(l)+f_R''-K(r)\right],
\end{equation}

\begin{equation}
\label{14a}
p_t^{eff}=\frac{1}{f_R}\left[p_t+\frac{r'(l)}{r(l)}f_R'-K(r)\right],
\end{equation}
Now, in the following subsections, we shall discuss various wormhole models by imposing various well-known forms of $f(R)$.

\subsection{Exponential $f(R)$ gravity}

\begin{widetext}
\begin{figure}[!htbp]
    \centering
    \includegraphics[width=\textwidth]{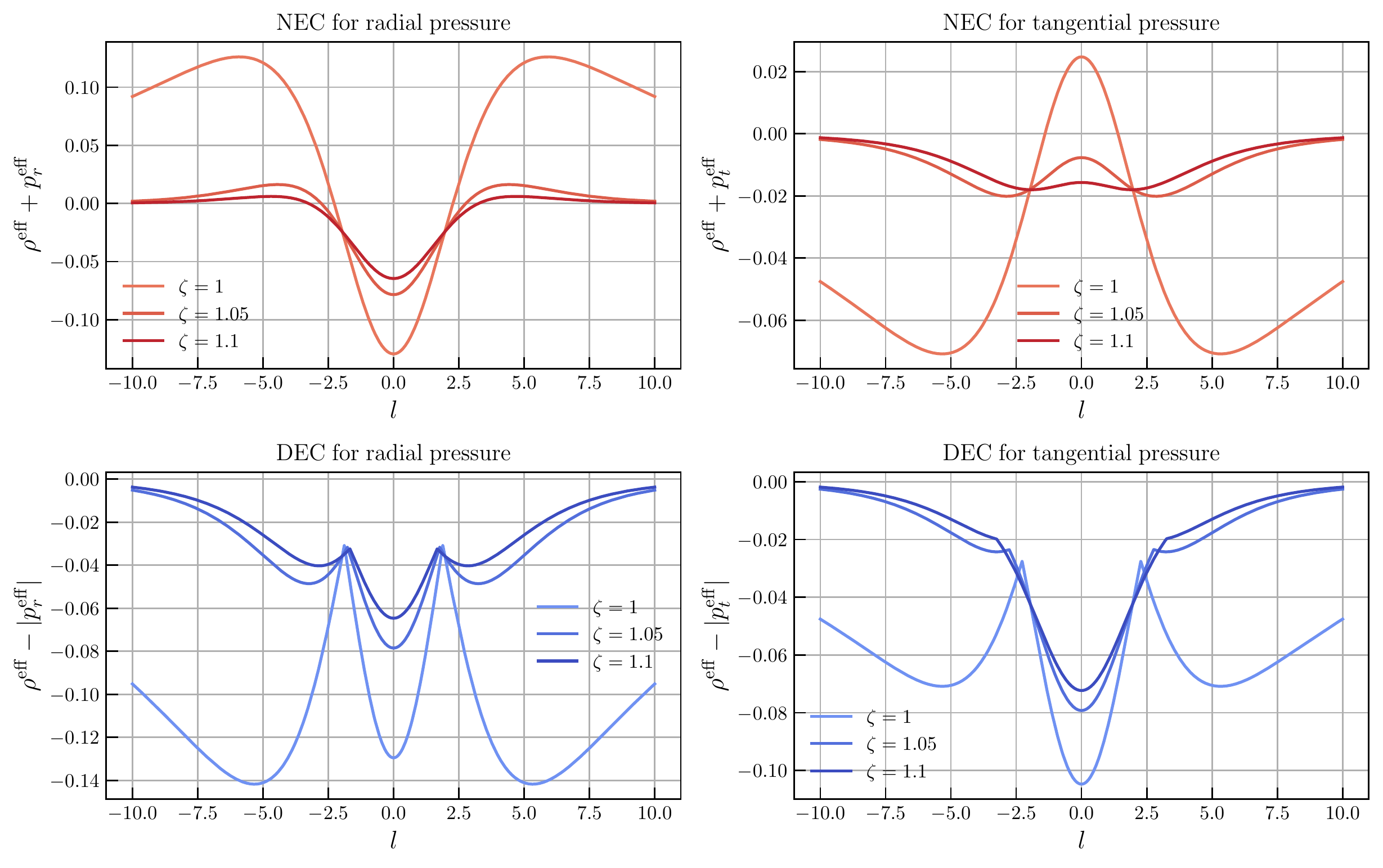}
    \caption{Null and Dominant energy conditions for GEB wormhole. To plot the results, we have considered that wormhole throat is equal to $b_0=5$, and cosmological constant is set to by unity (for simplicity). Furthermore, $m=2$ and $\alpha=0.5$, $\beta=0.1$}
    \label{fig:1}
\end{figure}
\end{widetext}

First physically viable model that we will consider in the current manuscript is namely exponential $f(R)$ gravity with the cosmological constant $\Lambda$ present \cite{ref65,ref66,ref67,ref68,ref69,ref70}:
\begin{equation}
    f(R)=R-2\Lambda\bigg[1-\exp \left(-\zeta \frac{R}{2\Lambda}\right)\bigg]
\end{equation}
Where $\zeta$ is positive free parameter. This kind of modified gravity could precisely describe the universe evolution with $z<10^4$. Thus, exponential MOG covers era of the recombination, matter dominated epoch and late-time accelerated expansion. To solve the field equations (\ref{12})-(\ref{14}), we firstly must assume the proper Equation of State. In the present paper we assume that the fluid is described by the barotropic EoS:
\begin{equation}
    p_r = \alpha \rho
    \label{eq:26}
\end{equation}
\begin{equation}
    p_t = \beta \rho
    \label{eq:27}
\end{equation}
where $\alpha\land\beta\in(0,1)$. With that assumption, from field equations it follows that energy density takes the form below:
\begin{widetext}
\begin{equation}
    \begin{gathered}
    \rho = \bigg[2 \exp \bigg(\frac{\zeta  \left(r'^2+2 r r''-1\right)}{\Lambda  r^2}\bigg) \bigg(r^4 \bigg(\zeta  \left(2 \zeta ^2 r'''^2+\zeta  \Lambda 
   r''' r'+\Lambda ^2 \left(r'^2-1\right)\right)-\Lambda ^2 \left(r'^2-1\right)\\
   \times \exp \bigg(-\frac{\zeta  \left(r'^2+2 r r''-1\right)}{\Lambda 
   r^2}\bigg)\bigg)+\Lambda  r^5 \bigg(2 \Lambda  r'' \bigg(\zeta -\exp \bigg(-\frac{\zeta  \left(r'^2+2 r r''-1\right)}{\Lambda 
   r^2}\bigg)\bigg)+\zeta ^2 r''''\bigg)+\zeta ^2 \Lambda  r^3 \left(1-3 r'^2\right) r''\\
   +\zeta ^2 r^2 r' \left(r'^2-1\right) \left(\Lambda 
   r'-4 \zeta  r'''\right)+2 \zeta ^3 r'^2 \left(r'^2-1\right)^2\bigg)\bigg]\bigg/\bigg[\Lambda ^2 (\alpha +2 \beta +3) r^6\bigg]
    \end{gathered}
\end{equation}
\end{widetext}
Then, using the equation above we firstly could derive effective energy density and anisotropic pressures and probe the aforementioned GEB wormhole energy conditions. We depicted EC's on the Figure (\ref{fig:1}) with the varying values of MOG free parameter $\zeta$ and fixed $m=2$, $\alpha=0.5$, $\beta=0.1$, $b_0=5$, and finally $\Lambda=1$. It is obvious that Null and Dominant Energy Conditions were violated everywhere for any value of $b_0$, $\alpha$ and $\beta$ and also $\mathrm{SEC}\approx\mathcal{O}(10^{-16})$. It is also interesting that NEC and DEC violation holds even for $m\neq2$.

\subsection{Hu-Sawicki $f(R)$ model}

Hu-Sawicki is the another type of the viable $f(R)$ gravity, for which the $f(R)$ function reads \cite{ref71}:
\begin{equation}
    f(R)=R-\frac{\lambda R_c (R/R_c)^{2n}}{1+(R/R_c)^{2n}}
\end{equation}
Where $\lambda$, $R_c$ and $n$ are dimensionless positive free parameters. This model includes the $\Lambda$CDM as a limit and could be seen as a late modification of the $\Lambda$CDM model \cite{ref72}. Then, by assuming barotropic EoS defined by the Equations (\ref{eq:26}) and (\ref{eq:27}) one could derive the energy density for Hu-Sawicki model of gravitation, but in the current article we will only show the numerical solutions because of the fact that expressions for energy density are too big.

\begin{widetext}

\begin{figure}[!htbp]
    \centering
    \includegraphics[width=\textwidth]{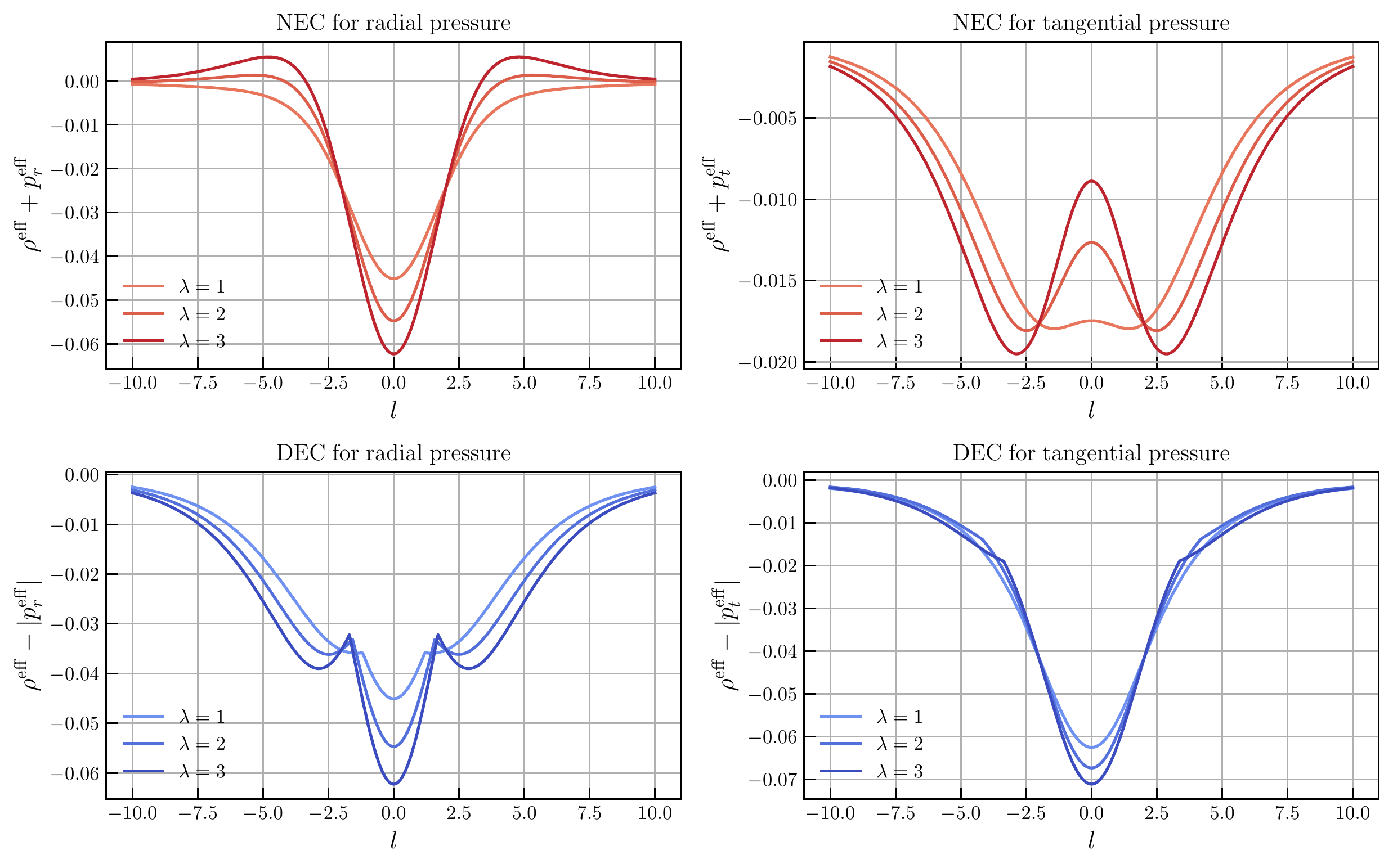}
    \caption{Null, Dominant and Strong energy conditions for GEB wormhole in the Hu-Sawicki gravity conjucture. To plot the results, we have considered that wormhole throat is equal to $b_0=5$, and also cosmological constant was set to unity. Furthermore, $m=2$ and $\alpha=0.5$, $\beta=0.1$, $R_c=n=1$}
    \label{fig:2}
\end{figure}
\end{widetext}

On the Figure (\ref{fig:2}) we plot the energy conditions for Hu-Sawicki $f(R)$ model with the varying values of MOG parameter $\lambda$ with positive bounds. As one may obviously notice from the plots, all of the aforementioned energy conditions except NEC were violated at the GEB wormhole throat (DEC is very similar to the exponential $f(R)$ gravity, thus $\mathrm{SEC}\approx \mathcal{O}(10^{-17})$). Moreover, situation does not differs even in that case, if we will vary other MOG parameters, such as $R_c$ and $n$. Also, even if we will consider more general case with $m\neq2$, energy conditions will be still violated.

\subsection{$\gamma$ gravity}

In this subsection we will investigate only one, last case of $f(R)$ gravity, the one which could provide the viable description of the universe using the gamma function, namely gamma gravity \cite{ref73}:
\begin{equation}
    f(R) = R- \frac{\alpha R_*}{n}\underbrace{\int ^{(R/R_*)^n}_0 k^{\frac{1}{n}-1}e^{-k}dk}_\text{$\gamma(1/n,(R/R_*)^n)$}
\end{equation}
Where $\alpha$, $n$, and $R_*$ are free positive MOG parameters. This kind of $f(R)$ gravity satisfies all of the stability and validity conditions, such as \cite{ref74}:
I) $f_{RR}>0$ (no tachyons), II)
$1+f_{R}>0$ (effective gravitational constant $G_{\mathrm{eff}}$ does not change sign, and thus no ghosts are present),

III)  
$\lim_{R\to\infty}f(R)/R=0$ and $\lim_{R\to\infty}f_{R}=0$ (GR is fully recovered at the early times), IV) $|f_{R}|$ is relatively small (solar and galactic scale constraints are satisfied).
This form of modified $f(R)$ gravity is statistically similar to the concordance $\Lambda\mathrm{CDM}$ model, but unlike it was for the Hu-Sawicki \cite{ref71} and Starobinsky \cite{ref75} modified gravity theories, $\gamma$ gravity does not include $\Lambda\mathrm{CDM}$ as a limit. 

\begin{widetext}

\begin{figure}[!htbp]
    \centering
    \includegraphics[width=\textwidth]{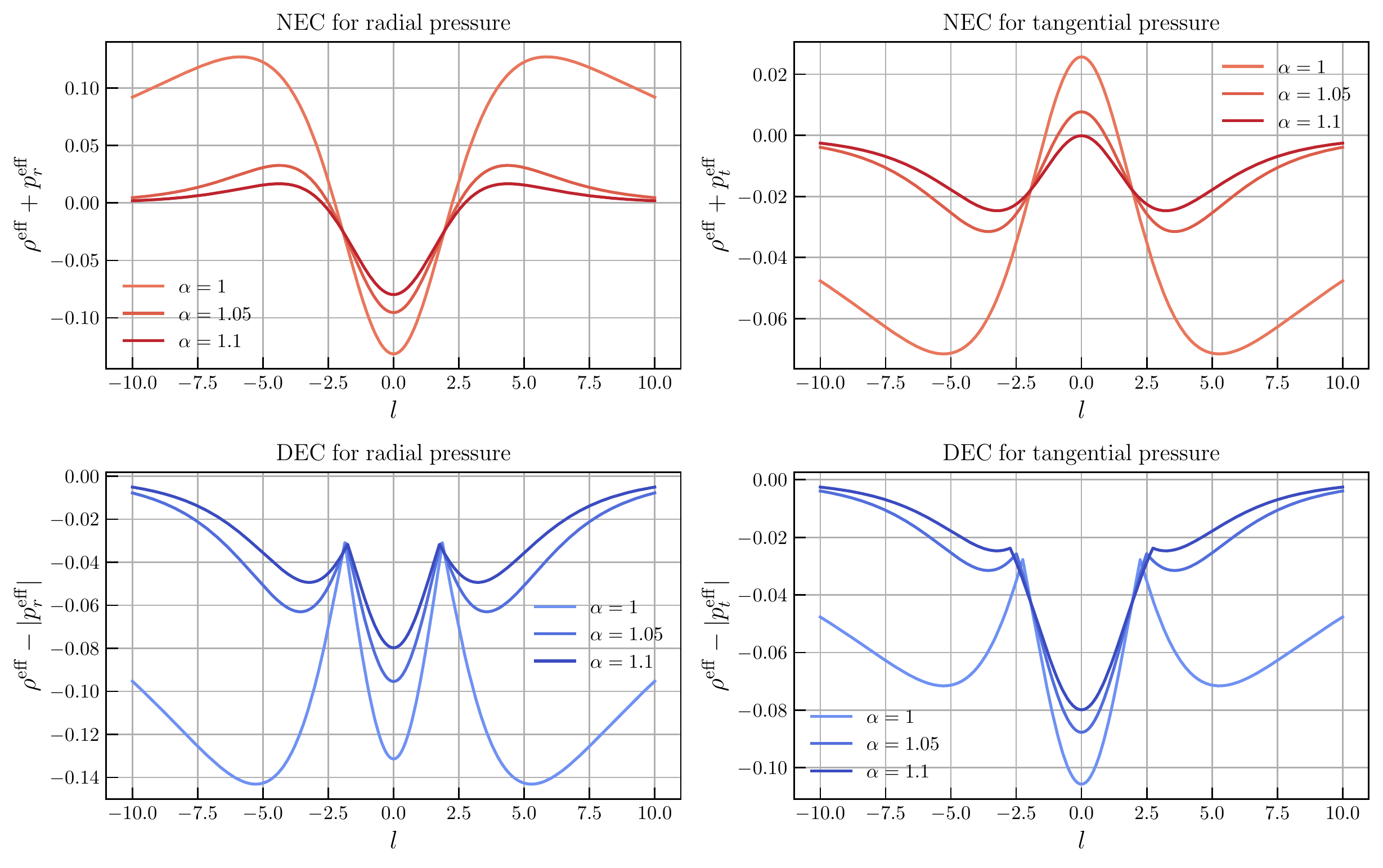}
    \caption{Null, Dominant and Strong energy conditions for GEB wormhole in the $\gamma$ $f(R)$ gravity. To plot the results, we have considered that wormhole throat is equal to $b_0=5$, cosmological constant is equal to $\Lambda=1$. Furthermore, $m=2$ and $\alpha=0.5$, $\beta=0.1$, $R_*=n=1$}
    \label{fig:3}
\end{figure}

\end{widetext}

As usual, we illustrate the null and dominated energy conditions at the Figure (\ref{fig:3}). We used $b_0=5$. For that case, Null Energy Condition as well as the Dominant Energy Condition was violated for every relatively small and positive value of $\alpha$ and $R_*$ (which is necessary condition judging by the observational constraints, presented in \cite{ref76}) if $n=1$. In turn, SEC has very small and positive values (so the situation is the same as in exponential and Hu-Sawicki gravities).

\section{Probing GEB wormhole stability through the MTOV}\label{sec4}

In the present subsection we will probe the stability of the perfect fluid matter that supports the generalised Ellis-Bronnikov wormhole interior.
For the propose of wormhole stability analysis one could use the equilibrium condition, which could be obtained from the well known generalized Tolman-Oppenheimer-Volkov (TOV) equation. The TOV equation is given below \cite{ref77,ref78}:
\begin{equation}
    \underbrace{\Phi'(\rho+p_r)}_\text{$F_G$}+\underbrace{\frac{dp_r}{dr}}_\text{$F_H$}+\underbrace{\frac{2}{r}(p_r-p_t)}_\text{$F_A$}+F_{E}=0.
    \label{eq:33}
\end{equation}
Where $F_G$ is the gravitational force, $F_H$ is the hydrodynamical one and $F_A$ is the contribution to the TOV of the fluid anisotropy. Finally, $F_E$ is the extra force, that arise because of the stress-energy tensor discontinuity ($\nabla^\mu T_{\mu\nu}\neq0$). Because we already considered only one case with ZTF GEB wormhole, gravitational force vanish. We need to explore the tortoise coordinate space, so we will apply the proper transformation:
\begin{equation}
dr^2=\left(1-\frac{b}{r}\right)dl^2.
\end{equation}
Keeping that fact in mind and by using the chain rule in the Leibniz notation, modified TOV equation reads:
\begin{multline}
\left(1-\frac{b(l)}{(b_0^2+l^m)^{1/m}}\right)^{-1/2}\frac{dp_r^{\mathrm{eff}}(l)}{dl}\\
 +\frac{2}{(b_0^2+l^m)^{1/m}}(p_r^{\mathrm{eff}}(l)-p_t^{\mathrm{eff}}(l))+F_{E}=0
 \label{eq:30}
\end{multline}
Where shape function is defined by the Equation (\ref{5}), but the shape function need to be defined in terms of tortoise coordinate using the expression $r(l)=(b_0^2+l^m)^{1/m}$.

\begin{widetext}
\begin{figure}[H]
    \centering
    \includegraphics[width=\textwidth]{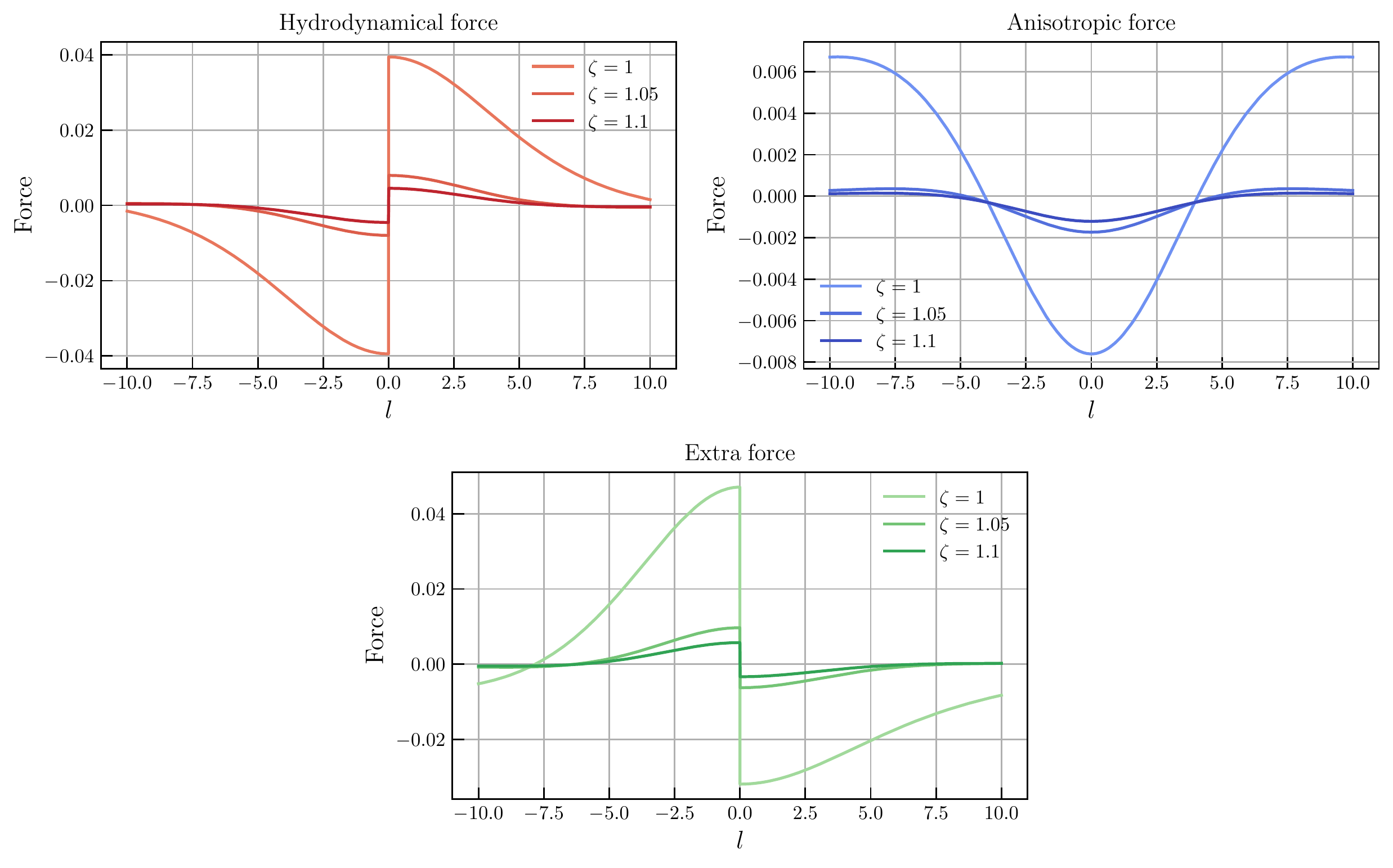}
    \caption{Forces that are present in the modified TOV equation for the exponential $f(R)$ gravity. As usual, we assume that $b_0=5$ and $\Lambda=1$, $m=2$, $\alpha=0.5$ and $\beta=0.1$.}
    \label{fig:5}
\end{figure}

\end{widetext}
On the Figure (\ref{fig:5}) we show the numerical solution of the Equation (\ref{eq:30}) for exponential $f(R)$ gravity.
In turn, we present the numerical results of MTOV forces evaluation for Hu-Sawicki $f(R)$ gravity on the Figure (\ref{fig:6}). In this kind of modified gravity, if we will vary the parameter $n$ (which represent the order of normalized scalar curvature $R/R_c$), as $n$ getting bigger, MTOV forces will be getting smaller equally. On the other hand, if we will vary other MOG free parameter $R_c$, as $R_c\to\infty$, $F\to 0$.
Finally, we show TOV forces for the last modified gravity of our consideration, namely gamma gravity on the Figure (\ref{fig:6a}). We plotted this forces for both linear and quadratic cases.
\begin{widetext}

\begin{figure}[!htbp]
    \centering
    \includegraphics[width=\textwidth]{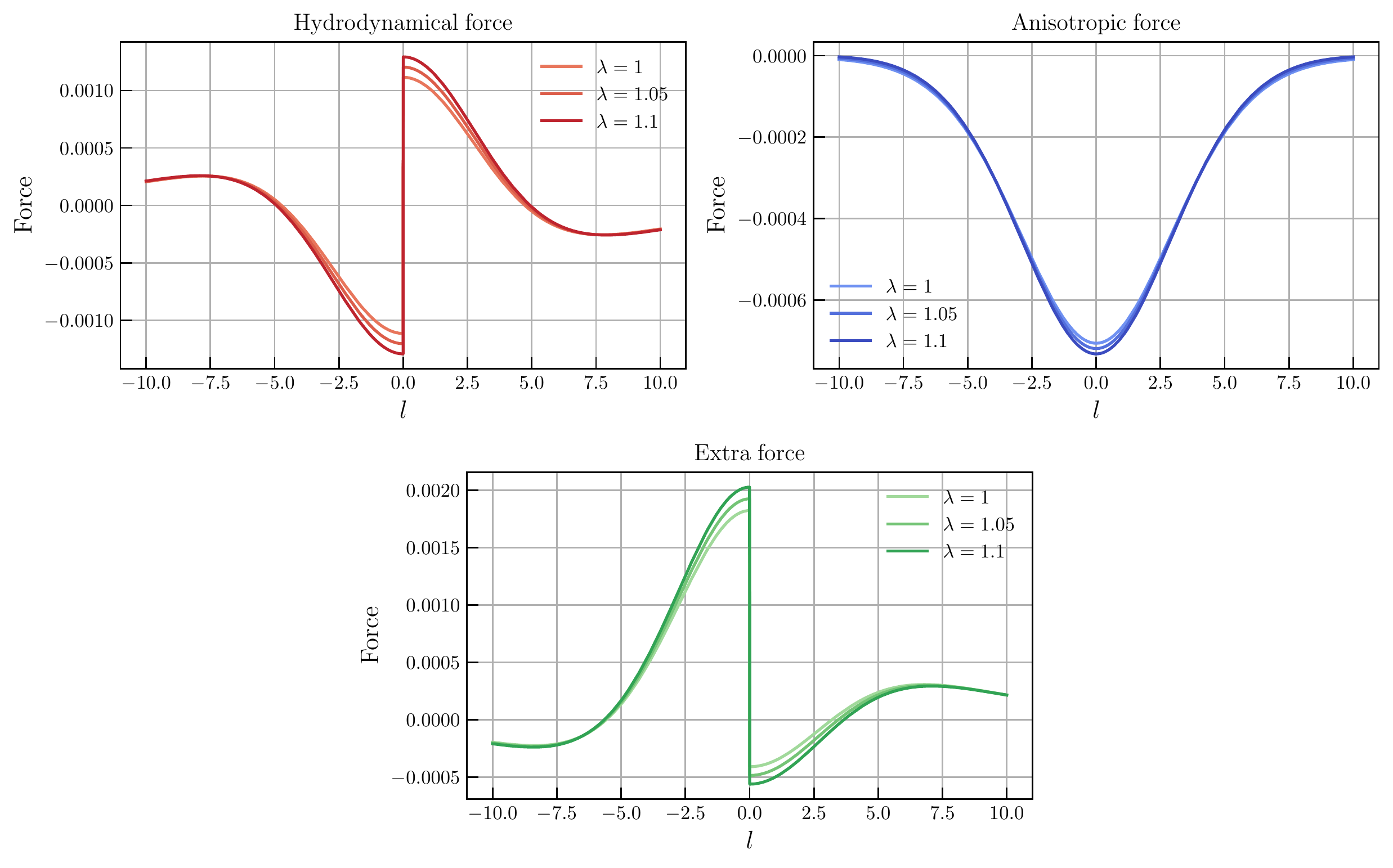}
    \caption{Forces that are present in the modified TOV equation for the Hu-Sawicki $f(R)$ gravity. As usual, we assume that $b_0=5$ and $\Lambda=1$, $m=2$, $\alpha=0.5$ and $\beta=0.1$, $R_c=1$}
    \label{fig:6}
\end{figure}



\begin{figure}[!htbp]
    \centering
    \includegraphics[width=\textwidth]{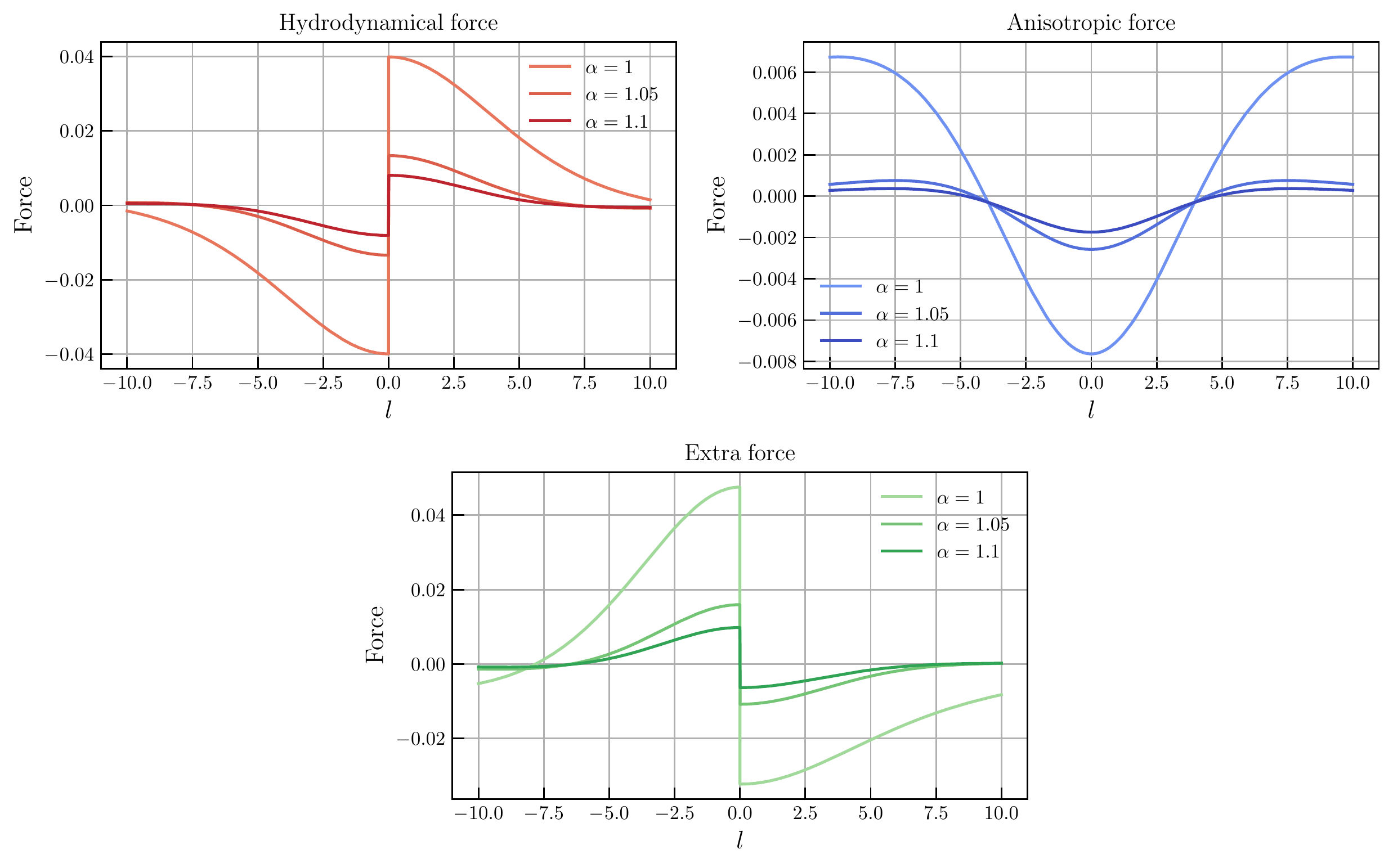}
    \caption{Forces that are present in the modified TOV equation for the gamma $f(R)$ gravity. As usual, we assume that $b_0=5$ and $\Lambda=1$, $m=2$, $\alpha=0.5$ and $\beta=0.1$. Moreover, to plot the figure we fixed the values of free MOG parameter $R_*$ to 1}
    \label{fig:6a}
\end{figure}

\end{widetext}

\section{Volume Integral Quantifier}\label{sec5}

\begin{widetext}

\begin{figure}[!htbp]
    \centering
    \includegraphics[width=\textwidth]{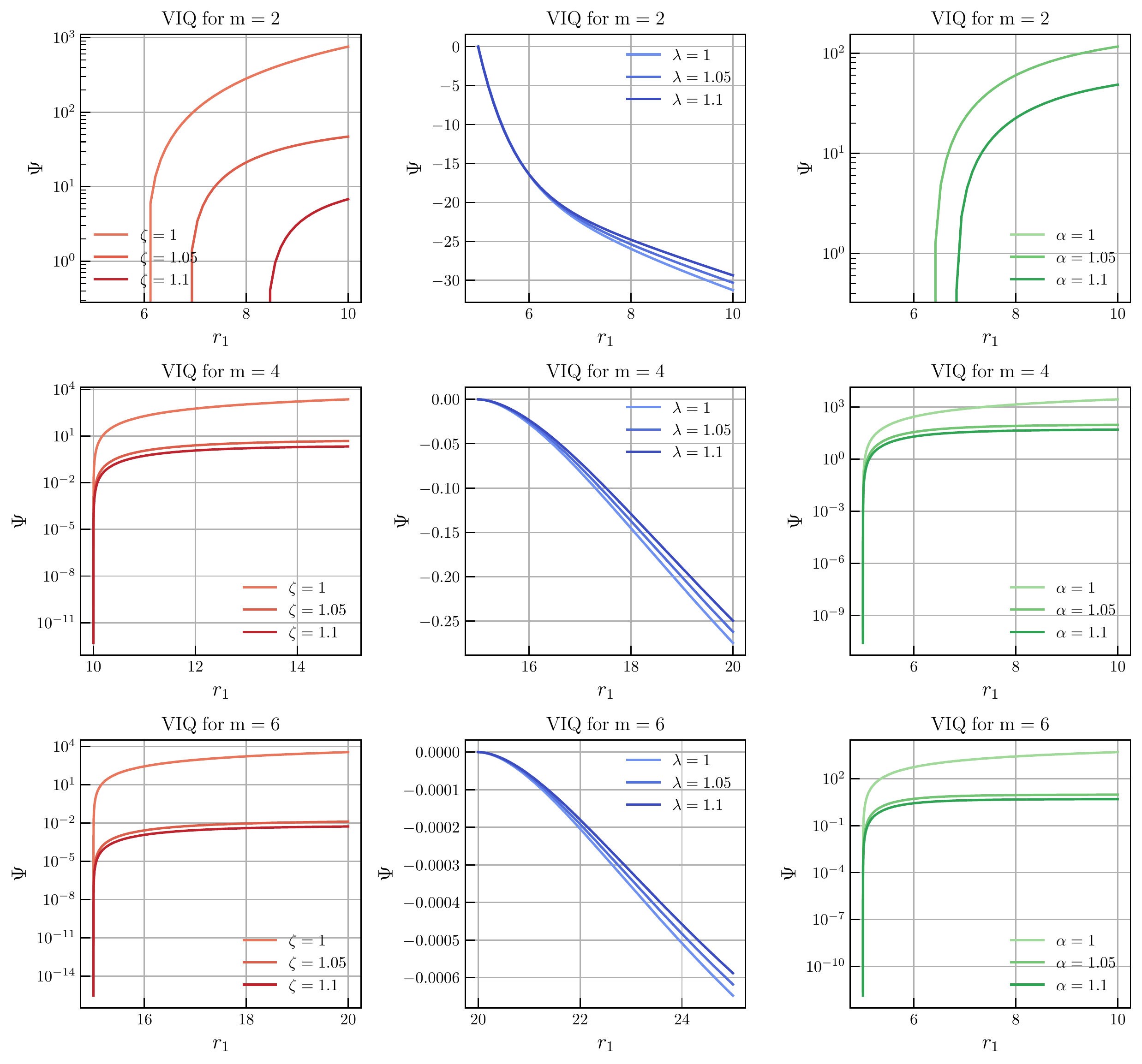}
    \caption{Volume Integral Quantifier for the exponential (\textit{first column}), Hu-Sawicki (\textit{second column}) and Gamma (\textit{third column}) gravities. Free MOG parameters take values that were assumed for energy conditions previously}
    \label{fig:8}
\end{figure}

\end{widetext}

Volume Integral Quantifier (here and further - just VIQ) could help us to quantify the total amount of Average Null Energy Condition (ANEC) violating matter (exotic matter) at some point of spacetime. Usually, VIQ for the spherically symmetric spacetime and anisotropic matter distribution takes the following form \cite{ref79}:
\begin{equation}
    \Psi = \int^\infty_{l_0}\int^\pi_0\int^{2\pi}_0[\rho+p_r]\sqrt{-g}drd\theta d\phi
    \label{eq:24}
\end{equation}
One could also rewrite VIQ (\ref{eq:24}) as the curvilinear integral over the volume $V$:
\begin{equation}
    \Psi = \oint [\rho+p_r]dV=2\int^\infty_{b_0}[\rho+p_r]4\pi r^2 dr
\end{equation}
But obviously we couldn't deal with the integration over infinite bounds (for the present solution, for a wormhole to be asymptotically flat, volume integral quantifier, integrated all over $r$ must have infinite values \cite{ref80}) and so we want to consider the VIQ with a cut-off of the stress energy tensor at some radius $r_1$:
\begin{equation}
    \Psi = 2\int^{r_1}_{b_0}[\rho+p_r]4\pi r^2 dr
    \label{eq:333}
\end{equation}
Then, we could finally rewrite Equation (\ref{eq:333}) in the terms of tortoise coordinate:
\begin{multline}
\Psi = \int^{\left(r_1^m-b_0^2\right)^{\frac{1}{m}}}_{\left(b_0^m-b_0^2\right)^{\frac{1}{m}}}[\rho^{\mathrm{eff}}(l)+p_r^{\mathrm{eff}}(l)]4\pi ((b_0^2+l^m)^{1/m})^2\\
\times \left(1-\frac{b}{(b_0^2+l^m)^{1/m}}\right)^{1/2}dl
\end{multline}
We routinely plot Volume Integral Quantifier for each modified gravity of our consideration on the Figure (\ref{fig:8}). It is easy to notice that Generalised Ellis-Bronnikov wormholes for $ m>1$ have exotic matter at the throat, but the total volume of ANCE violating matter could be minimised if we will assume that $\zeta\to\infty$, $\lambda\to\infty$ and $\alpha\to\infty$. But, as we know, contribution of gravity modification could not be very big, so maximally minimised GEB wormholes are not physically viable. 

\section{Total gravitational energy}\label{sec6}

\begin{widetext}

\begin{figure}[!htbp]
    \centering
    \includegraphics[width=\textwidth]{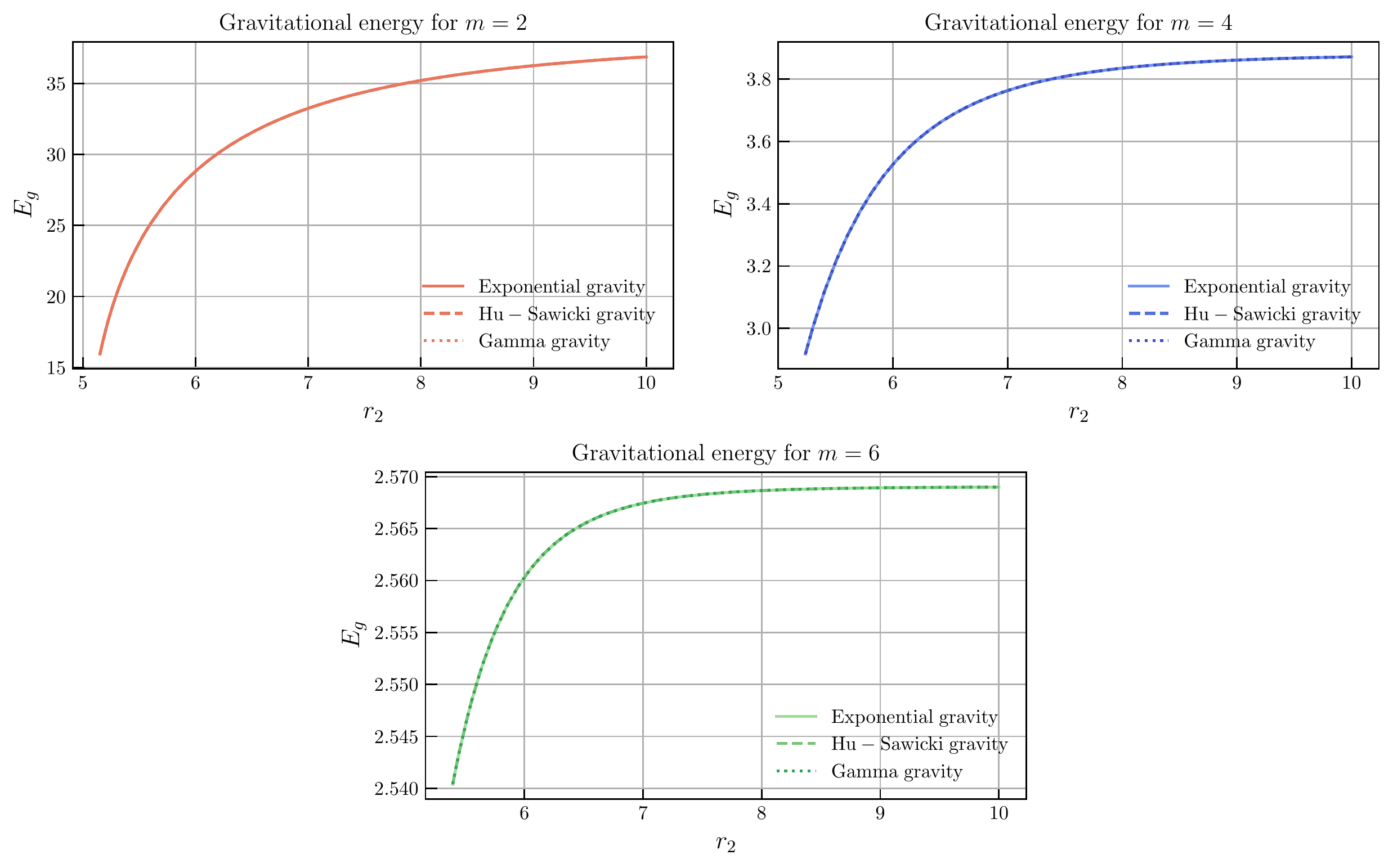}
    \caption{Total gravitational energy $E_{g}$ for the exponential, Hu-Sawicki and Gamma $f(R)$ gravity. As usual, we assume that $\Lambda=1$, $\alpha=0.5$ and $\beta=0.1$.}
    \label{fig:99}
\end{figure}

\end{widetext}

Total gravitational energy also could show us the behavior of the matter in the wormhole spacetime.  The total gravitational energy of a structure composed of normal baryonic matter is negative \cite{ref81}. For the first time total gravitational energy for any stationary spacetime (if we consider that black holes are absent) was discovered by the Lynden-Bell et al. \cite{ref82}. Total gravitational energy looked like $E_g=M-E_M$, where $M$ is the total mass and $E_M$ is the gravitational binding energy. But, it is easily to follow the work of \cite{ref83}, in which the explicit form of the $E_g$ were derived:
\begin{equation}
    E_g = M-E_M= \frac{1}{2}\int^{r_3}_{b_0}[1-\sqrt{g_{rr}}]\rho r^2 dr + \frac{b_0}{2}
    \label{eq:35}
\end{equation}
Here $b_0/2$ could be referred to as the effective gravitational mass. Then, we could rewrite explicit form of the total gravitational energy in terms of tortoise coordinate $r$:
\begin{multline}
    E_g = \frac{1}{2}\int^{\left(r_3^m-b_0^2\right)^{\frac{1}{m}}}_{\left(b_0^m-b_0^2\right)^{\frac{1}{m}}}\left[1-\left(1-\frac{b}{(b_0^2+l^m)^{1/m}}\right)^{-1/2}\right]\\
    \times\rho^{\mathrm{eff}} ((b_0^2+l^m)^{1/m})^2 \left(1-\frac{b}{(b_0^2+l^m)^{1/m}}\right)^{1/2}dl + \frac{b_0}{2}
    \label{eq:35}
\end{multline}
We plot the total gravitational energy for each gravity kind of our consideration with varying $m$ on the Figure (\ref{fig:99}). As one could easily notice, the biggest total gravitational energy has regular Ellis-Bronnikov wormhole, and if $m\to\infty$, then $E_g\to b_0/2$. Moreover, gravitational energy does not change if we will vary such MOG parameters as $\zeta$, $\lambda$ and $\alpha$, $R_c$ and $R_*$, so it is invariant under the change of gravitation formalism.

\section{Concluding remarks}\label{sec7}

Wormholes are fascinating objects in the spacetime structure, and its existence appears in theoretical physics as a solution to Einstein field equations. But these are yet to be confirmed through observations/experiments in the background of unified field theory. In classical general relativity, violation of NEC is a basic necessary condition to a static traversable wormhole. Despite this, NEC and WEC can be avoided for a time-dependent wormhole solution in specific regions and for a particular interval of time at the throat \cite{ref84,ref85,ref86,ref87}. Moreover, in the alternative theories of gravity to GR, by modifying the Einstein-Hilbert action, one may impose in principle that the energy-momentum tensor looping the wormhole validates the NEC. However, later NEC is necessarily violating in the context of the effective energy-momentum tensor. For instance, in the case of braneworld wormhole solutions, the matter contents on the brane satisfy the NEC, whereas effective energy-momentum tensor violates it later \cite{ref43,ref44,ref45,ref46}.
It is worthy to note here that the WEC can be satisfied depending on the parameters of the gravitational theory \cite{ref41}.

In this manuscript, we have explored the possibilities of the wormhole geometries in the framework of $f(R)$ gravity theory. We considered the GEB spacetime for our study and derived the modified motion equations for the test particle. Further, we examined three wormhole geometries by taking various well-established $f(R)$ models such as exponential $f(R)$ gravity, Hu-Sawicki $f(R)$ model, and $\gamma$ model. These models are well-known for their successful description of late-time cosmic acceleration and their concordance with the $\Lambda$CDM model. Moreover, the energy conditions are tested for the wormhole models. And, it is observed that NEC is violated near the throat of the wormhole for all the models. These results indicates the presence of exotic matter, which helps the traveler to pass through the wormhole throat freely.

As a matter of completeness, we tested some physical properties such as stability through MTOV, volume integral quantifier (VIQ), and total gravitational energy for the wormhole models. The stability of the wormhole models is examined by the hydrodynamical force $F_H$, an-isotropic force $F_A$, and extra force $F_E$. And their combining results satisfied the equilibrium condition. From VIQ profiles, it is seen that wormholes have exotic matter at the throat, but the total volume of ANCE violating matter could be minimised if we will assume that $\zeta\to \infty$,\, $\lambda\rightarrow \infty$, and $\alpha\rightarrow\infty$. But, as we know, the contribution of gravity modification could not be enormous, so maximally minimized GEB wormholes are not physically viable. From the total gravitational energy profiles, one can easily observe that the biggest total gravitational energy has a regular Ellis-Bronnikov wormhole, and if $m\to\infty$, then $E_g\to b_0/2$. Moreover, gravitational energy does not change if we will vary such MOG parameters as $\zeta$, $\lambda$ and $\alpha$, $R_c$ and $R_*$, so it is invariant under the change of gravitation formalism.

These above-discussed results allowed us to verify different wormhole geometries in the context of $f(R)$ gravity theories with the GEB line element, lightening a new possibility of wormhole geometry. Besides this, it would be interesting to study GEB wormhole geometries by taking account of the coupling of $f(R)$ with the inflanton fields. We intend to explore some of these studies in the near future and hope to report on them.

\section*{Acknowledgments}

S.M. acknowledges Department of Science \& Technology (DST), Govt. of India, New Delhi, for awarding Senior Research Fellowship (File No. DST/INSPIRE Fellowship/ 2018/IF180676). PKS acknowledges National Board for Higher Mathematics (NBHM) under Department of Atomic Energy (DAE), Govt. of India for financial support to carry out the Research project No.: 02011/3/2022 NBHM(R.P.)/R\&D II/2152 Dt.14.02.2022.  We are very much grateful to the honorable referee and to the editor for the illuminating suggestions that have significantly improved our work in terms of research quality, and presentation.

\end{document}